\documentclass[sigconf]{acmart}

\usepackage[english]{babel}
\usepackage{blindtext}
\usepackage[utf8]{inputenc}
\usepackage{subfigure}
\usepackage{array}

\renewcommand\footnotetextcopyrightpermission[1]{} 
\setcopyright{none}

\acmDOI{}

\acmISBN{}

\acmPrice{}

\begin{document}


\title{On Machine Learning DoS Attack Identification from Cloud Computing Telemetry}


 \author{João Henrique Corrêa, Patrick Marques Ciarelli,\\Moises R. N. Ribeiro, Rodolfo da Silva Villaca}
 \orcid{0000-0002-8124-8985}
 \affiliation{%
   \institution{Federal University of Espírito Santo}
   \streetaddress{Av. Fernando Ferrari, 514 - Goiabeiras - Vitória - Espírito Santo - Brasil}
   \city{Vitória - Espírito Santo - Brasil} 
 }
 \email{jhenrique@gmail.com, patrick.ciarelli@ufes.br, moises.ribeiro@ufes.br, rodolfo.villaca@ufes.br}

\renewcommand{\shortauthors}{João H. G. M. Corrêa et al.}

\begin{abstract}
    The detection of Denial of Service (DoS) attacks remains a challenge for the cloud environment, affecting a massive number of services and applications hosted by such virtualized infrastructures. Typically, in the literature, the detection of DoS attacks is performed solely by analyzing the traffic of packets in the network. This work advocates for the use of telemetry from the cloud to detect DoS attacks using Machine Learning algorithms. Our hypothesis is based on  richness of such native data collection services, with metrics from both physical and virtual hosts. Our preliminary results demonstrate that DoS can be identified accurately with k-Nearest Neighbors (kNN) and decision tree (CART).
\end{abstract}

\maketitle

\section{Introduction}

Denial of Service (DoS) attacks are a pressing problem in computer networks, and it becomes an even more complex problem in cloud computing environments. According to the Verisign report of 2018, these kinds of attacks directly impact business, financial, information technology and telecommunications services \cite{verisign-ddos}.

The migration of traditional services and applications to centralized cloud environments amplifies their vulnerability to disruption in relation to the impact of DoS attacks performed over conventional architectures. Thus, there is an urgent need to perform the identification and characterization of DoS traffic in cloud environments. In general, there are three ways to perform traffic classification: i) identification using TCP/UDP port numbers; ii) inspecting the contents of the network packets; and, finally, iii) applying Machine Learning (ML) techniques to network data source \cite{ml-sdn-2016}. This paper advances the latter, and  proposes the use of ML algorithms in order to distinguish malicious traffic from legitimate traffic from clients in a cloud environment. ML algorithms have been already used for traffic identification and classification \cite{dl-survey}, including malicious ones, to spot the occurrence of a DoS attack.

According to a recent survey \cite{Boutaba2018}, the traditional data source used by ML algorithms are traces of traffic in the network (packets). It is well-known that DoS attacks systemically affects the usage of cloud computing resources. Different from traditional approached based on traffic traces, this work proposed the use of the telemetry from the cloud (such as resources usage from physical and virtual hosts) as data source for ML algorithms. 

Large scale monitoring traffic in conventional networks usually involves costly and complex architectures, probe packets and other artifices. In contrast, clouds have native telemetry, i.e., data collection services. Metrics from both physical and virtual hosts. Moreover, such rich data set can be used with little or no overhead to the cloud. Therefore, the research hypothesis of this paper is that it is possible to improve identification and accuracy of DoS attacks using the information from the usage of computing resources from physical and virtual hosts allocated to the applications and services.

\section{Evaluation}

As an initial scenario of this proposal, we monitor the usage of computing resources of a cloud under the TCP SYN Flood attack. This attack has been chosen because it is simple to perform and reproduce. An experimental setup has been created with an OpenStack cloud (Rocky version), containing a web server using Apache2, a virtual machine simulating legitimate web clients with Siege \footnote{https://www.joedog.org/siege-home/}. Finally, another host performs the role of malicious users and generates the SYN Flood Attack with hping3\footnote{https://linux.die.net/man/8/hping3}.

\begin{figure*}[ht!]
\centering
\hfill
\subfigure[CPU\label{fig:memory}]{
\includegraphics[width=0.22\textwidth,trim={0cm 0cm 1.5cm 1.3cm},clip]{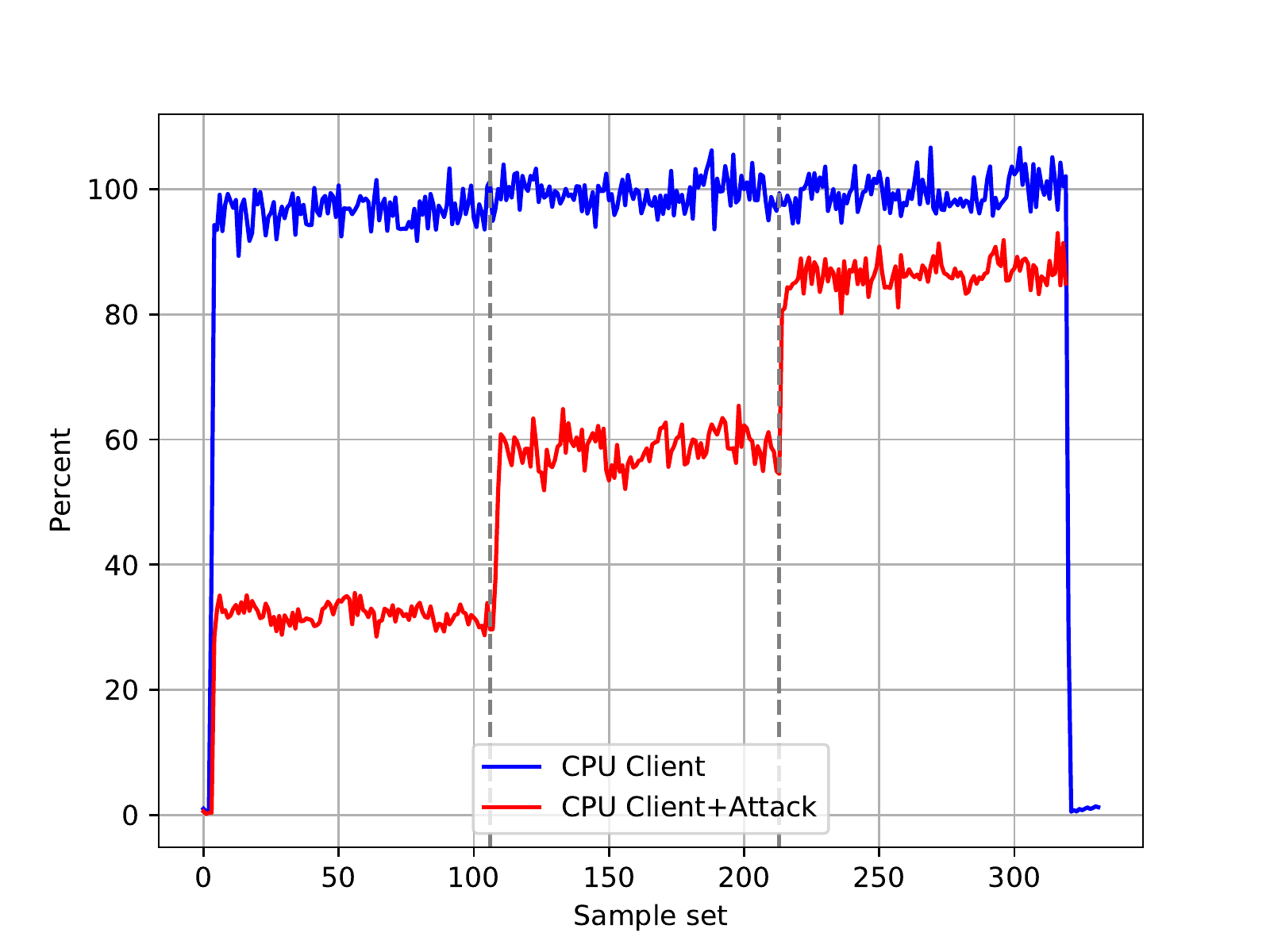}}
\hfill
\subfigure[Memory\label{fig:memory}]{
\includegraphics[width=0.22\textwidth,trim={0cm 0cm 1.5cm 1.3cm},clip]{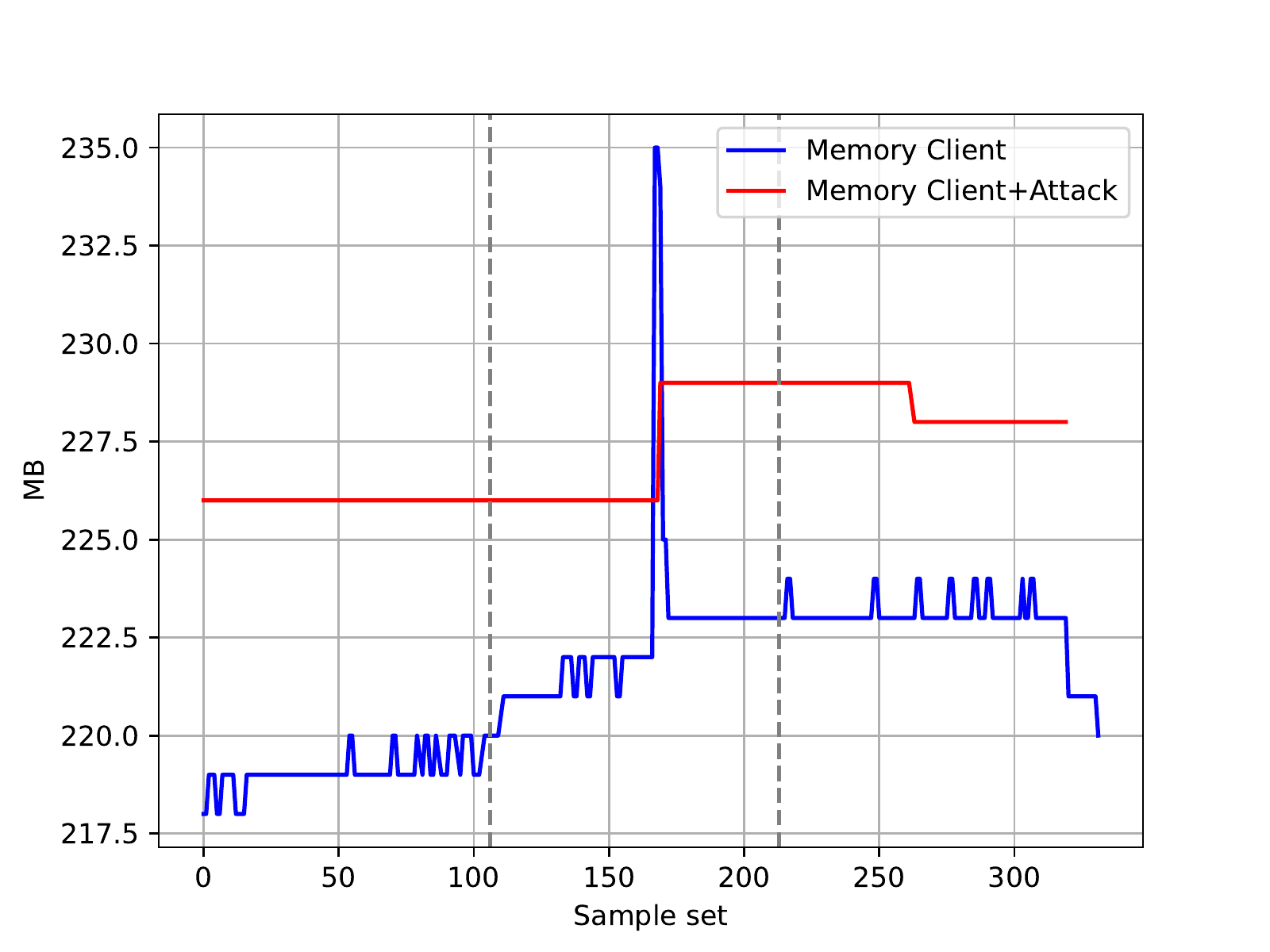}}
\hfill
\subfigure[HD Request read\label{fig:hd-read}]{
\includegraphics[width=0.22\textwidth,trim={0cm 0cm 1.5cm 1.3cm},clip]{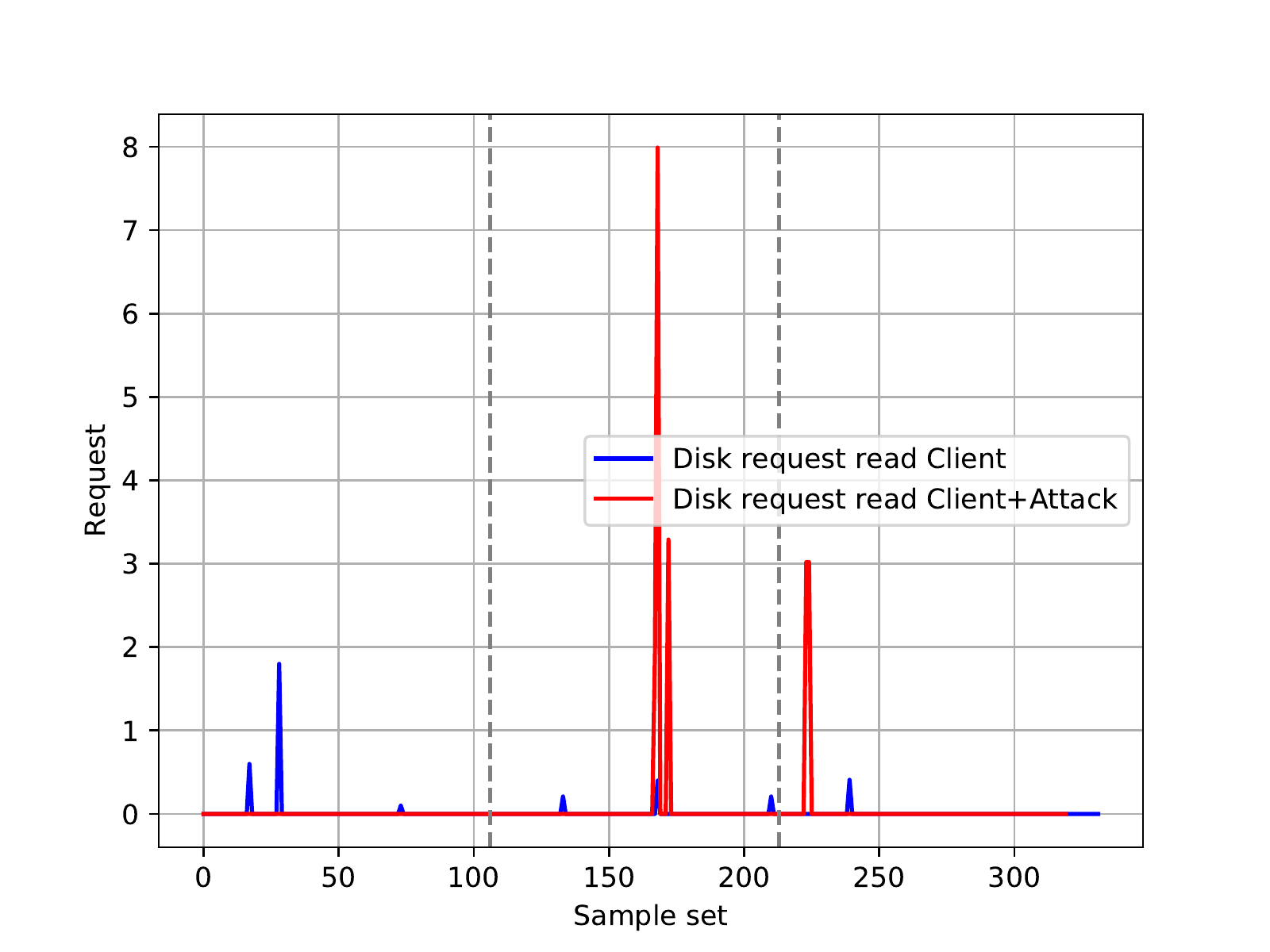}}
\hfill
\subfigure[HD Request write\label{fig:hd-write}]{
\includegraphics[width=0.22\textwidth,trim={0cm 0cm 1.5cm 1.3cm},clip]{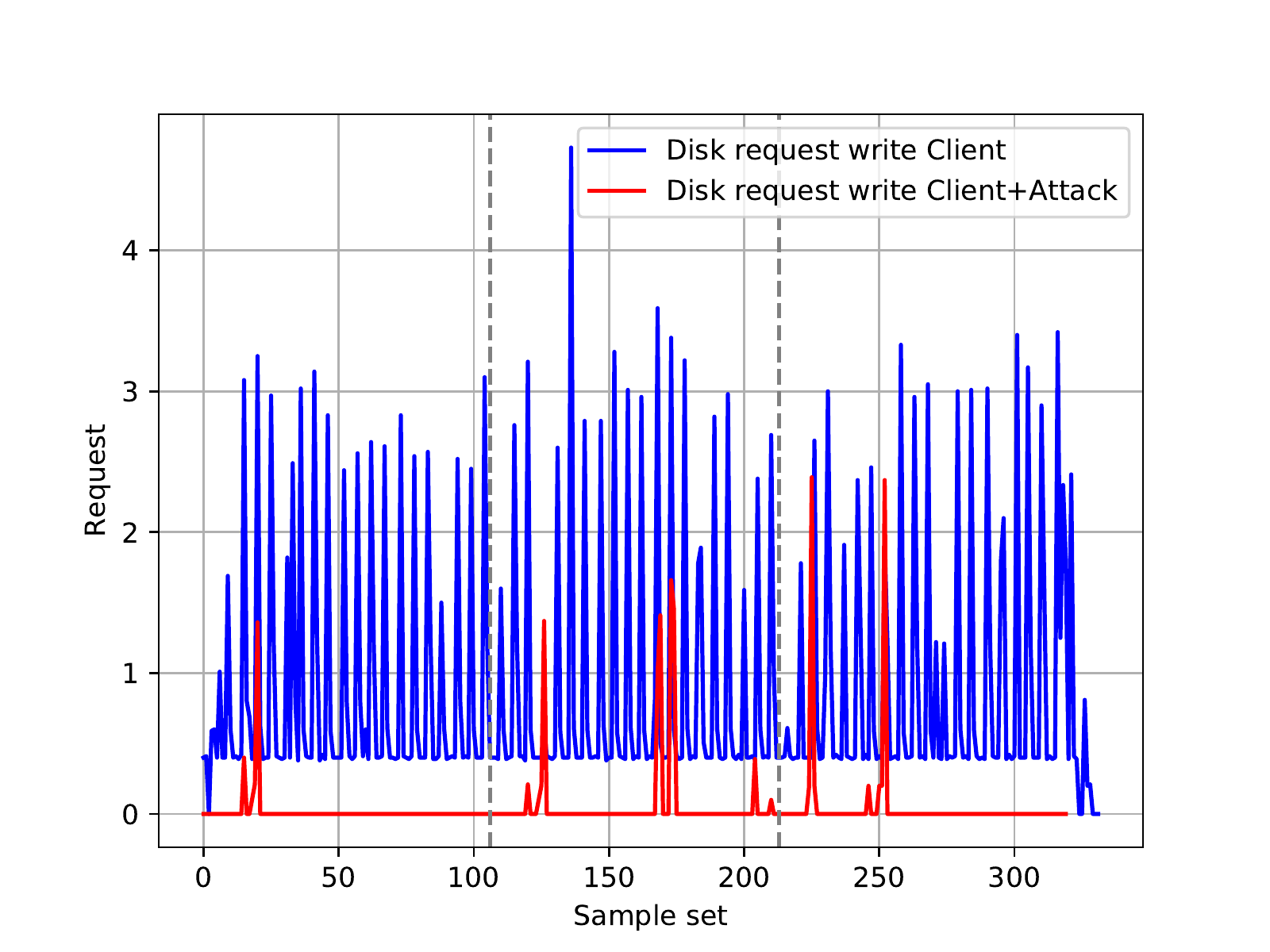}}
\hfill
\vspace{-0.2cm}
\subfigure[Interface Bytes in\label{fig:bytes-in}]{
\includegraphics[width=0.22\textwidth,trim={0cm 0cm 1.5cm 1.3cm},clip]{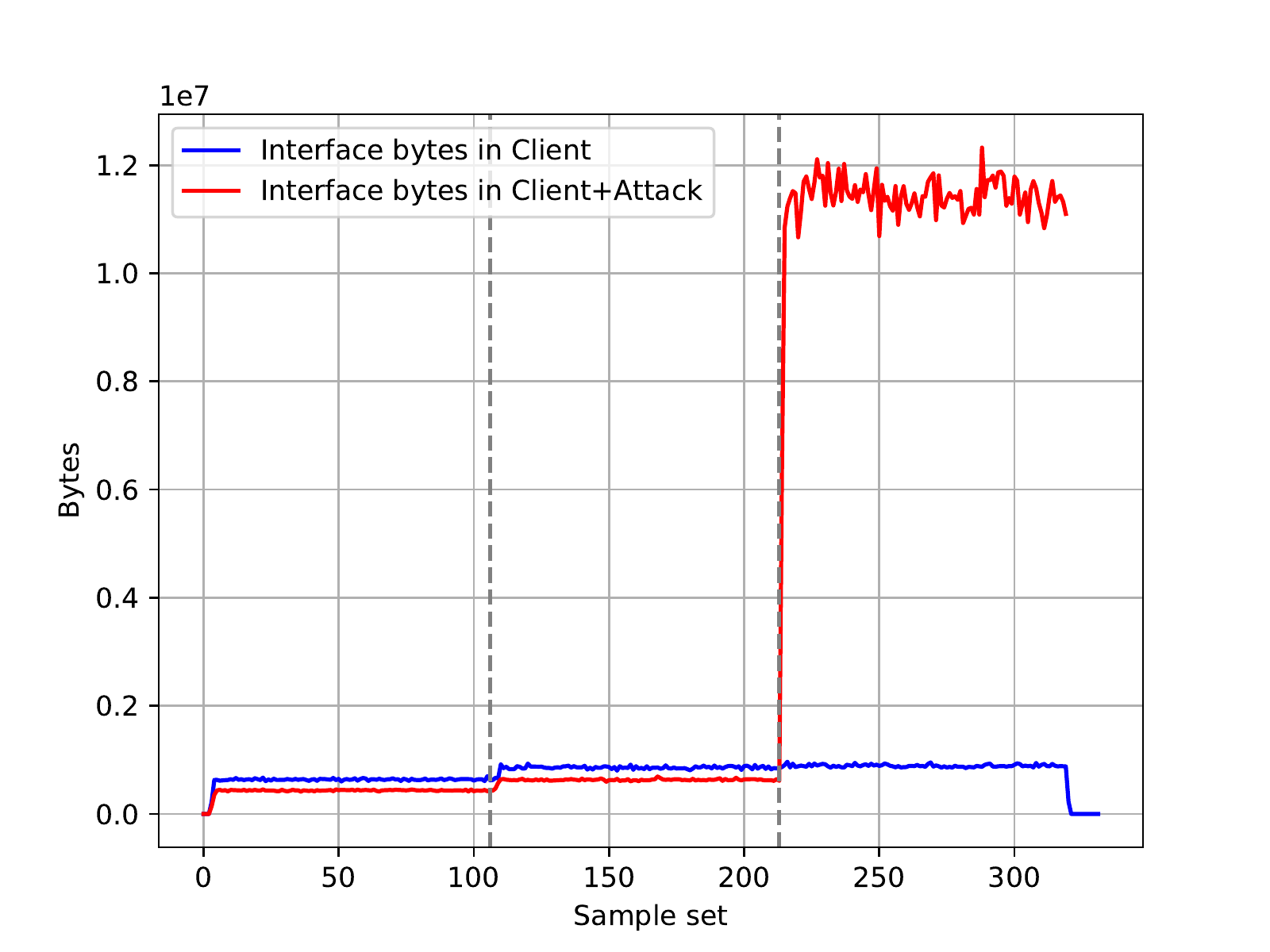}}
\hfill
\subfigure[Interface Bytes out\label{fig:bytes-out}]{
\includegraphics[width=0.22\textwidth,trim={0cm 0cm 1.5cm 1.3cm},clip]{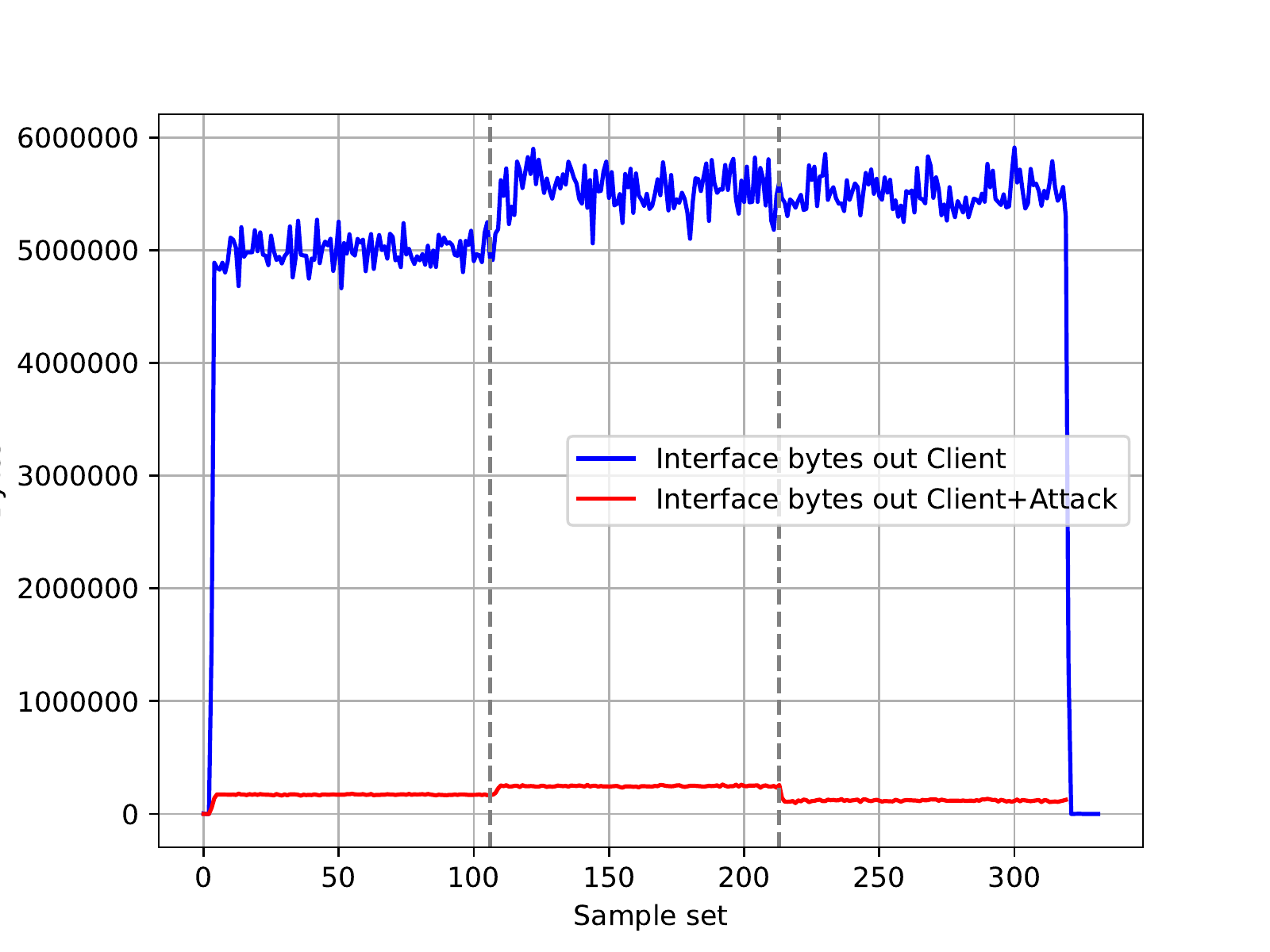}}
\hfill
\subfigure[Interface packets in\label{fig:packets-in}]{
\includegraphics[width=0.22\textwidth,trim={0cm 0cm 1.5cm 1.3cm},clip]{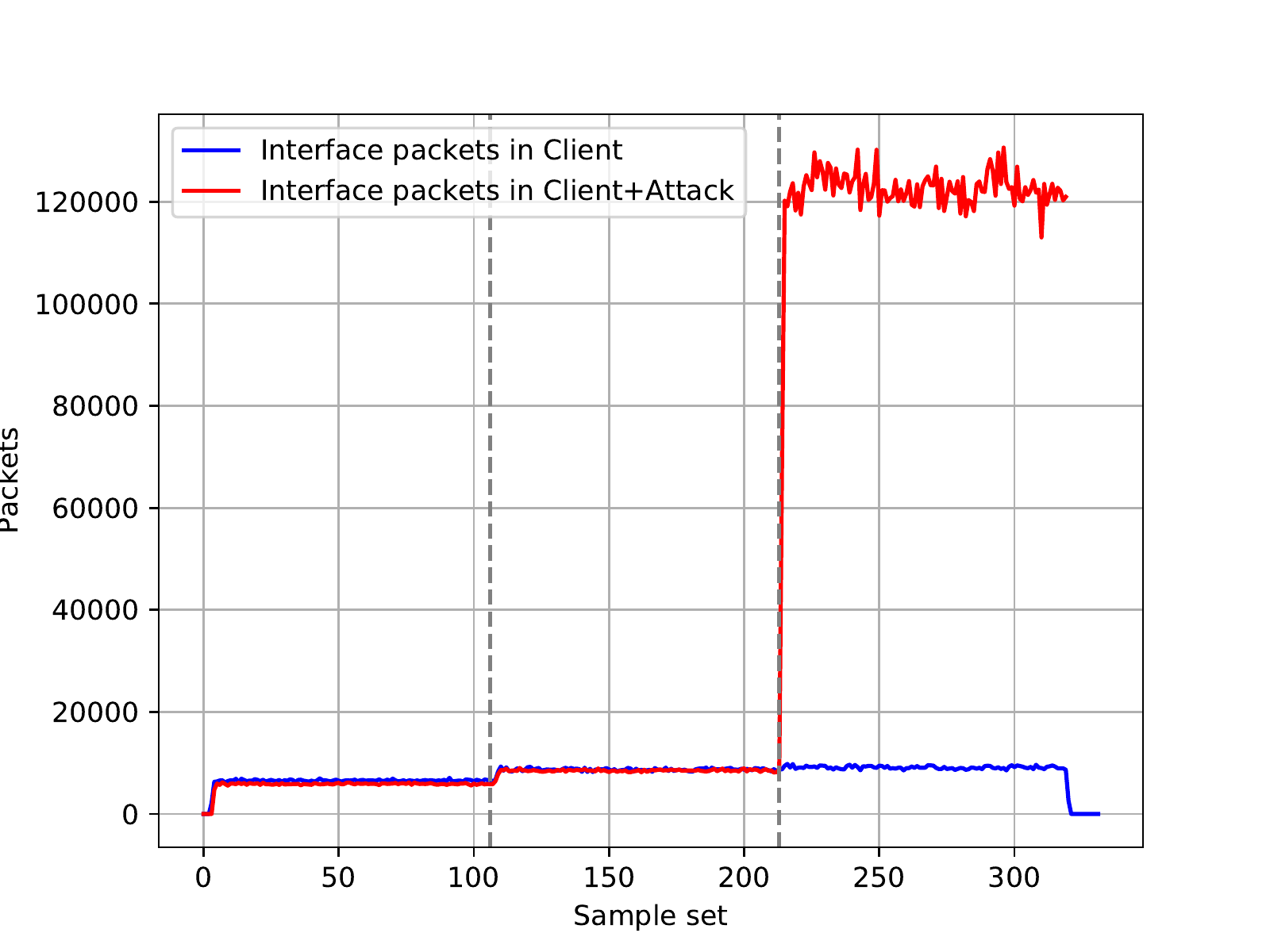}}
\hfill
\subfigure[Interface packets out\label{fig:packets-out}]{
\includegraphics[width=0.22\textwidth,trim={0cm 0cm 1.5cm 1.3cm},clip]{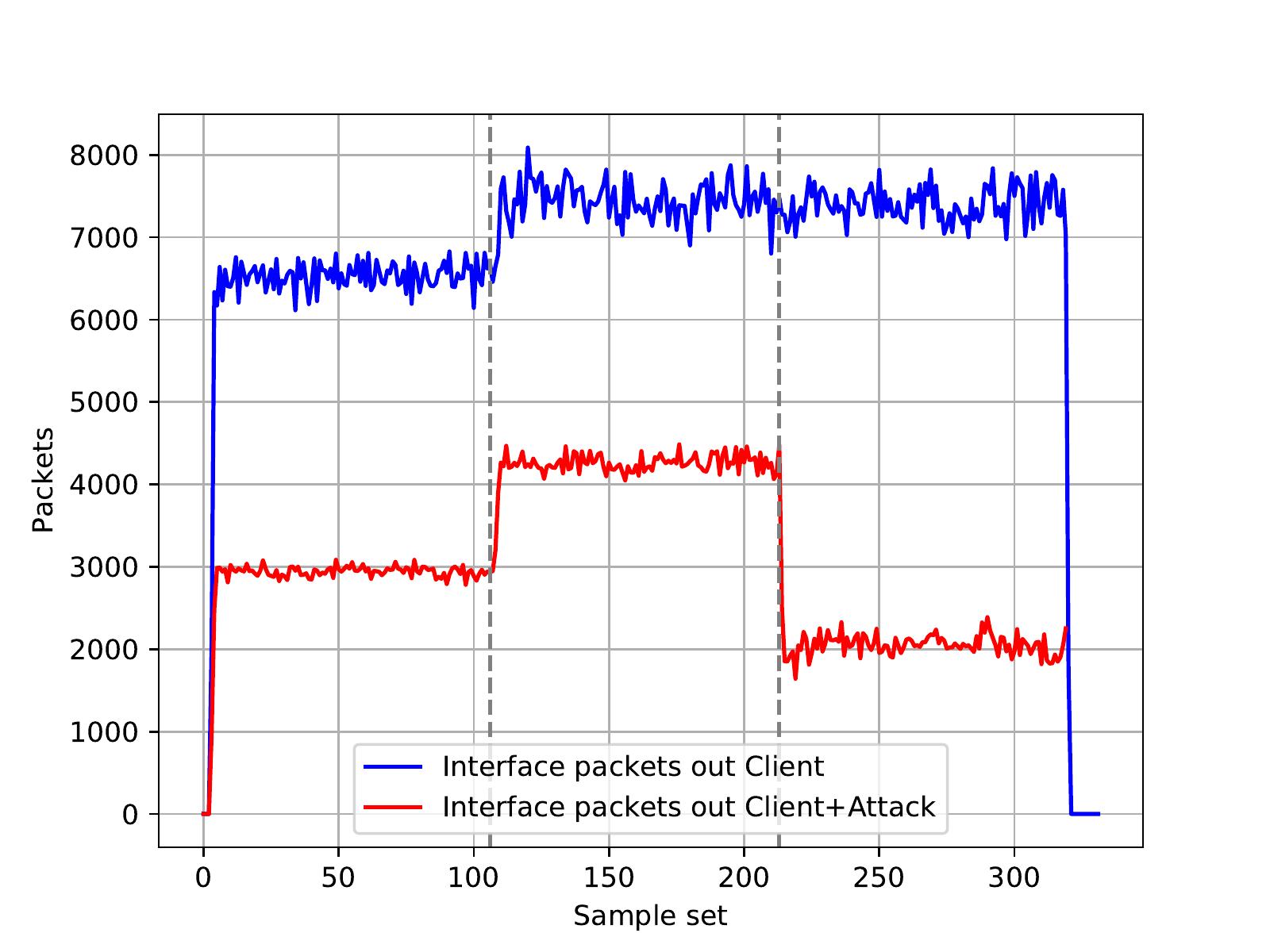}}
\hfill
\vspace{-0.4cm}
\caption{Comparison between the metrics in both scenarios.}
\label{fig:gra}
\vspace{-0.3cm}
\end{figure*}

Two test scenarios have been performed. In the first test there is only legitimate clients, generating traffic at from users so that the web server can hold it. In the second test, however, along with legitimate clients, SYN Flood attack is executed. During both tests, the following metrics are monitored by the OpenStack cloud: CPU and Memory usage, number of disk read requests, number of disk write requests, rate of incoming and outgoing Bytes in each network interface, rate of incoming and outgoing packets in each network interface.

In both tests, the total duration is 30 minutes. There are attack phases with different intensities, which last 10 minutes each. There are 10s gaps at beginning and at ending parts of our trace with no legitimate clients' nor attacker's traffic. Metrics are collected every 5 seconds performing a total of 360 samples. The attack is divided into three phases of 10 minutes each. In the first one, the attack happens at a small rate i.e.,  at every 300 milliseconds. In the second phase, 10 minutes after, the malicious attacks happens every 250 milliseconds. Finally, in the last phase, the attack is generated at the maximum rate that the attacking host can perform.

Given the measurements of the two experiments, a Principal Component Analysis (PCA) was used to verify which metrics are relevant in the case of a TCP SYN Flood attack. A sample comparison of these metrics is shown in Figure~\ref{fig:gra}. In all the results, the blue lines represent the tests performed only in the client scenario and the red lines in the client and attacker scenario. According to these results the following metrics were chosen: CPU usage, number of disk write requests, rate of incoming and outgoing Bytes, rate of incoming and outgoing packets. 


After the selection of the most important characteristics, the traffic labeling in the data is performed. These samples are used as a training dataset for two different ML algorithms: k-Nearest Neighbors (kNN) and a decision tree (CART). These simple algorithms were chosen for the sake of simplicity to obtain preliminary results in order to verify our hypothesis regarding ML exploiting could telemetry. The data from this dataset has been labeled in two classes: "attack" and "no attack". 

In addition, another experiment was carried out, with a duration of 120 minutes, alternating between moments of traffic with only legitimate clients, only attacker and a combination between both legitimate clients and attackers. With the collected metrics of this experiment, the trained kNN and CART tries to identify attack periods. 

The Table~\ref{tab:ml} presents the results (in percentage) of the Accuracy, Precision, Recall and F1-Score metrics obtained in the dataset using the kNN and CART algorithms, being the first line for the kNN and the last line for the decision tree algorithm (CART). In these preliminary tests a high accuracy was verified, with low incidence of false positives and false negatives, mainly in the kNN algorithm.

\begin{table}
\vspace{-0.2cm}
\begin{center}
\caption{Result of ML algorithms}
\vspace{-0.4cm}
\label{tab:ml}
\small
\begin{tabular}{|c|c|c|c|c|}
\hline
  ML Algorithms & Accuracy & Precision & Recall & F1-Score \\ 
\hline
kNN & 99.30 & 99.33 & 99.31 & 99.31 \\ 
\hline
CART & 91.07 & 95.53 & 91.08 & 92.38 \\
\hline
\end{tabular}
\end{center}
\vspace{-0.5cm}
\end{table}

\section{Conclusion}

Given the promising results, the proposal to use data from the resources usage from cloud computing environment is a feasible hypothesis. 

As future work, we aim at performing the characterization of other types of DoS attacks, as well as combinations of these types. We intend not only to detect that an attack is occurring (anomaly detection), but also, through ML algorithms, to perform (real-time) the identification of what kind of attack is going on. Another future work is to explore different ML algorithms, including the use of deep learning.

\bibliographystyle{ACM-Reference-Format}
\bibliography{reference}

\end{document}